\documentclass[5p]{elsarticle}
\usepackage{pdfpages}

\usepackage{graphicx}
\usepackage{epstopdf}

\begin{document}


\title{The threshold displacement energy of buckminsterfullerene C$_{60}$ and formation of the endohedral defect fullerene He@C$_{59}$}

\author[su]{Mark~H.~Stockett\corref{mycorrespondingauthor}}
\cortext[mycorrespondingauthor]{Corresponding author}
\ead{Mark.Stockett@fysik.su.se}

\author[su]{Michael~Wolf}
\author[su,ui]{Michael~Gatchell}
\author[su]{Henning~T.~Schmidt}
\author[su]{Henning~Zettergren}
\author[su]{Henrik~Cederquist}

\address[su]{Department of Physics, Stockholm University, Roslagstullsbacken 21, SE-106 91 Stockholm, Sweden}
\address[ui]{Institut f\"{u}r Ionenphysik und Angewandte Physik, Universit\"{a}t Innsbruck, Technikerstr.\ 25, A-6020 Innsbruck, Austria}

\begin{abstract}
We have measured the threshold center-of-mass kinetic energy for knocking out a single carbon atom from C$_{60}^-$ in collisions with He. Combining this experimental result with classical molecular dynamics simulations, we determine a semi-empirical value of 24.1$\pm$0.5~eV for the threshold displacement energy, the energy needed to remove a single carbon atom from the C$_{60}$ cage. We report the first observation of an endohedral complex with an odd number of carbon atoms, He@C$_{59}^-$, and discuss its formation and decay mechanisms.
\end{abstract}


\maketitle

\section{Introduction}

Irradiation by electrons, ions, or atoms may displace individual atoms from their original positions in solids \cite{McKenna2016}, small aggregates of matter \cite{Zettergren2013}, or free molecules \cite{Stockett2014a}. Such defects can for example be used to tailor nanostructured materials with new intriguing functionalities \cite{Banhart2010,Krasheninnikov2010,Kotakoski2014}, and serve as reactive sites for efficient molecular growth processes inside molecular clusters \cite{Zettergren2013,Seitz2013,Wang2014}, for example in different planetary atmospheres. Introductions of defects may also be a limiting factor in high-resolution transmission electron microscopy, as material modifications during image capture may yield images that do not represent the original sample material \cite{Egerton1999,Dmytrenko2006}. In addition, collisions with H and He atoms leading to atom displacement are thought to be important mechanisms for destruction of large molecules in the interstellar medium \cite{micelotta2010,postma14}. The key intrinsic (projectile independent) target property quantifying the effect of these types of radiation damage is the so-called threshold displacement energy, $T_d$, the minimum energy transfer to a single atom required to permanently displace it from its initial position \cite{McKenna2016}. 

The threshold displacement energy is distinct from related quantities such as the dissociation energy or the vacancy energy in that it includes the energy barrier between the parent and product states. While independent of projectile, $T_d$ depends somewhat on the angle between the momentum imparted to the primary knock-on atom and the inter-atomic bonds \cite{cui94,Gatchell2016}. Nevertheless, typical values of $T_d$ are widely used to model radiation damage across wide ranges of energy. In recent breakthrough experiments \cite{Meyer2012,Meyer2013}, the threshold displacement energy for single-layer graphene was reported to be $T_d=23.6$~eV. Similar values for $T_d$ have also been deduced for gas-phase Polycyclic Aromatic Hydrocarbon (PAH) molecules \cite{Stockett2015}. 

Here, we determine the C$_{60} \rightarrow$ C$_{59}$+C threshold displacement energy for isolated C$_{60}$, which has a perfectly symmetric icosahedral structure where the C atoms have a different hybridization than graphene or PAHs. With all atoms equivalent, C$_{60}$ is the ideal prototype for displacement studies of molecules. Previously reported values of $T_d$ for C$_{60}$ from electron microscopy experiments and from Molecular Dynamics simulations have spanned a wide range from 7.6-15.7~eV \cite{Fueller1996} and 29.1~eV \cite{cui94}, respectively. Models of radiation damage to fullerenes have generally assumed a value in the middle of this range, around 15~eV, equivalent to the vacancy energy \cite{Parilis1994,Larsen1999,Zettergren2013}. 

In addition to the threshold displacement energy for C$_{60}$, we also report the first observation of the endohedral defect fullerene complex He@C$_{59}^-$ formed in collisions where the He is captured following C displacement. Endohedral complexes B@C$_{2n}$, where an atom or molecule B is trapped inside the fullerene cage, has been of interest since the dawn of the fullerene era, both from pure and applied perspectives \cite{Weiske1991,Sprang1994,Campbell1998,Komatsu2005,Kurotobi2011}. The present observation of He@C$_{59}^-$ provides insight into the formation and stability of endohedral complexes with odd numbers of carbon atoms.

In typical fullerene fragmentation experiments, energy deposited through interactions with \textit{e.g.} photons, electrons, or fast ions is converted into internal vibrational energy. For C$_{60}$, where C$_{60}\rightarrow$ C$_{58}$+C$_2$ is the lowest energy dissociation channel \cite{tomita01,TOMITA2003}, this leads to the well-known statistical product distribution dominated by fragments with even numbers of carbon atoms, C$_{60-2n}$, $n = 1,2,...$. In collision experiments like those presented here, products with odd numbers of C atoms like C$_{59}$ have occasionally been observed \cite{Weiske1991,Larsen1999,tomita01,Gatchell2014c}. These products are fingerprints of non-statistical fragmentation, where carbon atoms are displaced in billiard-ball like collisions \cite{Stockett2014a}. This process takes place on timescales that are too short (sub-femtoseconds) for local excitations to distribute over the whole molecular system. The exceptionally low yield of C$_{59}$ relative to statistical fragmentation products has so far precluded systematic experimental studies of this mechanism.

Here, we have developed a refined approach to determine displacement energies for free molecules in collisions with particles that improves upon earlier such methods in an important way, namely by eliminating the (usually dominant) contribution of statistical fragmentation from the product distribution. This is achieved by colliding He with C$_{60}^-$ ions and measuring the threshold behavior for the formation of \textit{negatively charged} fragments \textit{e.g.} C$_{59}^-$ and C$_{58}^-$. Because the electron affinity of C$_{60}$ (2.664~eV \cite{Stochkel2013}) is much lower than any of the dissociation energies of the system (which are $>10$~eV \cite{tomita01,TOMITA2003}, see also Tables \ref{tab_eb} and \ref{tab_diss} below), any trajectories depositing enough energy to induce statistical unimolecular dissociation most likely lead to electron loss and thus do not contribute to the negative ion product spectrum. In this way, we select those trajectories where essentially all the excitation energy is transferred to the primary knock-on atom, and as little as possible to the other atoms in the C$_{60}$ cage or to the electronic degrees of freedom. By eliminating the major source of background from the measurement, greater sensitivity to minute cross-sections for non-statistical fragmentation is achieved, without which the present results for C$_{60}$ would not be possible. We combine experimental measurements of the threshold center-of-mass energy for non-statistical fragmentation of C$_{60}^-$ in collisions with He with classical Molecular Dynamics (MD) simulations of the knockout process and a statistical model of electron loss from C$_{60}^-$ and C$_{59}^-$ to determine the threshold displacement energy $T_d$. Because the extra electron in C$_{60}^-$ has only a small effect on the binding between C atoms, this result applies to neutral as well as anionic C$_{60}$.

\section{Methods}

\subsection{Experiments}
\label{sec_exp}

Experiments were performed using the Electrospray Ion Source Laboratory (EISLAB) accelerator mass spectrometer, which has been described previously \cite{Stockett2014,deRuette2018}. Continuous beams of C$_{60}^-$ were produced by means of ElectroSpray Ionization, with tetrathiafulvalene (TTF) added as an electron donor \cite{tomita02} to a 1:1 mixture of methanol and dichloromethane containing fullerite. In the ion source, the C$_{60}^-$ ions undergo many low-energy collisions and are assumed to equilibrate to roughly room temperature. Following production of the ions by ESI, an ion funnel was used to collect and focus the ions. Two octupole ion guides transported the ions through two stages of differential pumping. A quadrupole mass filter was used to mass-select C$_{60}^-$ ions, which were then accelerated to 3--15~keV and passed through a 4 cm long collision cell containing the target gas (He). This C$_{60}^-$ kinetic energy range corresponds to center-of-mass energies, $E_{CM}$, of 20--80~eV for collisions with He in the cell. An Einzel lens and two pairs of electrostatic deflector plates served as a large-angular-acceptance kinetic-energy-per-charge analyzer after the gas cell. Intact C$_{60}^-$ and daughter anions were detected on a 40 mm multi-channel plate (MCP) with a position-sensitive resistive anode. The deflection voltage combined with the position on the MCP (along the direction of deflection) of each detected anion was used to calculate its kinetic energy per charge \cite{deRuette2018}. 

Total destruction cross sections (\textit{i.e.} the sum of the cross sections for all processes -- fragmentation and/or electron-loss or ionization processes -- that change the mass-to-charge ratio of C$_{60}^-$ in collisions with He under the present experimental conditions) were determined by measuring the attenuation of the parent C$_{60}^-$ beam as a function of target gas density. The attenuation is fit to a single exponential decay

\begin{equation}
\Gamma (p)=\Gamma_0e^{-p\sigma l/k_BT}
\end{equation}

\noindent where $\Gamma (p)$ is the parent C$_{60}^-$ count rate as a function of the He pressure $p$ in the gas cell, $\Gamma_0$ is the count rate at $p=0$, $\sigma$ is the total C$_{60}^-$ destruction cross section to be determined, $l$ is the length of the gas cell (4~cm), and $k_B$ is Boltzmann's constant. The temperature $T$ is 300~K.

In order to determine the absolute cross section for the formation of a specific daughter ion such as C$_{59}^-$, we distribute the total cross section $\sigma$ according to the corresponding number of product ions $N_i(p)$ relative to the number of intact parent ions $N_{par}(p)$ detected in the same spectrum: 

\begin{equation}
\sigma_i=\sigma\frac{N_i(p)}{N_{par}(p)(e^{p\sigma l/k_BT}-1)},
\end{equation}

\noindent where $N_{par}(p)(e^{p\sigma l/k_BT}-1)$ corresponds to the total number of parent ions destroyed as this spectrum was recorded.

\subsection{Calculations}
\label{sec_sims}

\begin{figure}
\includegraphics[width=0.98\columnwidth]{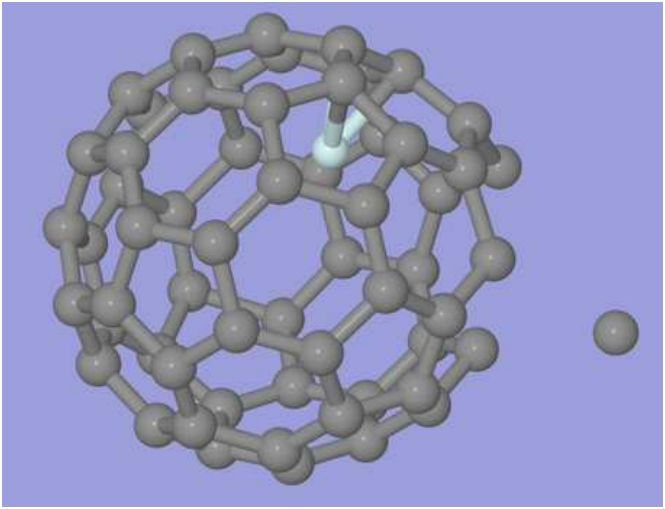}
\caption{Image from a video of a MD trajecotry leading to He@C$_{59}^-$. The C atom (black) at the right of the frame has been knocked out by the He atom (blue), which has been captured by the C$_{59}$ cage. The full video is available in the electronic version of this article.}
\label{fig_video}
\end{figure}

Classical molecular dynamics simulations of collisions between C$_{60}$ and He were carried out using the LAMMPS software package \cite{Plimpton:1995aa}. We used the reactive Tersoff potential \cite{PhysRevB.37.6991,PhysRevB.39.5566} to describe the bonds between C atoms and the Ziegler-Biersack-Littmark (ZBL) potential \cite{zbl_pot_book} for the He-C$_{60}$ interaction. For each collision energy 10,000 randomly oriented trajectories were simulated (with random orientations of the C$_{60}$ molecule) and followed for 500\,fs with a time step of $5\times 10^{-18}$\,s. One frame from a trajectory leading the formation of the He@C$_{59}^-$ reaction product is shown in Figure \ref{fig_video}. The full video of this trajectory is available in the electronic version of this article.

Density Functional Theory (DFT) was used to calculate electron detachment and dissociation energies of C$_{60}^-$ and C$_{59}^-$. These calculations were performed using Gaussian09 \cite{Frisch:2009aa_long} at the B3LYP/6-31+G(d) level. Following the removal of a single carbon atom from C$_{60}$, there are two closed-cage structures of C$_{59}$ to which the system may relax. These differ in the bonding between the three C atoms which were adjacent to the knocked out C; the naming here is adopted from Ref. \cite{Wang2014}. The $ho$-C$_{59}^-$ isomer, in which the cage-closing leads to the formation of a hexagonal (\textit{h}) and an octagonal (\textit{o}) ring at the vacancy cite, is 1~eV lower in energy than $pn$-C$_{59}^-$, where a pentagonal (\textit{p}) and a nonagonal (\textit{n}) ring are formed. For C$_{58}^-$, the lowest energy neutral isomer from Ref. \cite{chen07}, labeled $C_{3v}$-C$_{58}$, was used as a starting point for the calculations. 

\section{Results}

\subsection{Stability of C$_{59}^-$}

\begin{table}
\begin{tabular}{|c|ccc|}
\hline
Charge & C$_{60}$ & $ho$-C$_{59}$ & $pn$-C$_{59}$ \\
\hline
-1 & 2.59 & 3.56 & 3.37 \\
0 & 7.50 & 6.69 & 6.93 \\
\hline
\end{tabular}
\caption{Electron binding energies (in eV) for fullerene anions and neutrals calculated with DFT at the B3LYP/6-31+G(d) level. See text for isomer naming convention.}
\label{tab_eb}
\end{table}

Electron binding energies for both neutral and anionic C$_{60}$ and C$_{59}$ (both \textit{ho} and \textit{pn} isomers) are presented in Table \ref{tab_eb}. The present value for the electron affinity of C$_{60}$, 2.59~eV, is close to measured values (\textit{e.g.} 2.664~eV \cite{Stochkel2013}). Because of the dangling bonds, the electron affinity for both isomers of C$_{59}$ are much higher at 3.56~eV for $ho$-C$_{59}$ and 3.37~eV for $pn$-C$_{59}$. 

\begin{table}
\begin{tabular}{|c|ccc|}
\hline
 & C$_{60}\rightarrow$ & C$_{60}\rightarrow$ & C$_{59}\rightarrow$ \\
Charge & C$_{59}+$C & C$_{58}+$C$_2$ & C$_{58}+$C \\
\hline
-1 & 11.6 & 10.1 & 4.54 \\
0 & 12.6 & & \\
+1 & 11.8 & & \\
\hline
\end{tabular}
\caption{Dissociation energies (in eV) for lowest energy isomers of C$_{59}$ and C$_{58}$, $ho$-C$_{59}$ and $C_{3v}$-C$_{58}$, in various charge states calculated with DFT at the B3LYP/6-31+G(d) level.}
\label{tab_diss}
\end{table}

DFT-calculated dissociation energies for the main decay channels are given in Table \ref{tab_diss}. Our value for the C$_{60}^-\rightarrow$ C$_{58}^-$+C$_2$ dissociation energy of 10.1~eV is similar to the 10-11~eV measured for neutral and cationic C$_{60}$ \cite{TOMITA2003,tomita01}. The dissociation energy for C$_{60}\rightarrow$ C$_{59}$+C, which we calculate for anionic, neutral and cationic species, is around 12~eV which is significantly higher than for $C_2$-loss from C$_{60}$ in all cases and in agreement with previous calculations (for neutrals and cations \cite{Wang2014}). The dissociation energy calculated, however, does not consider any energy barrier between the the initial and final states. Given that C$_{59}$ is almost never observed in fullerene fragmentation experiments, even when employing excitation energies far in excess of all dissociation energies, it is possible that much higher barriers are present for C$_{60}\rightarrow$ C$_{59}$+C than for C$_{60}\rightarrow$ C$_{58}$+C$_2$ fragmentation processes. In contrast to the dissociation energy, the threshold displacement energy $T_d$ reflects the barriers for the C$_{60}\rightarrow$ C$_{59}$+C process. 

Our calculated dissociation energy for C$_{59}^-\rightarrow $ C$_{58}^-$+C is rather low at 4.54~eV and close to the previously calculated 5.4~eV for C$_{59}^+$ \cite{Wang2014}. This is also likely to contribute to the low observed yield of C$_{59}$ in experiments -- the products are fragile and may rather easily undergo delayed fragmentation processes. Internal energies upwards of 40~eV are known to be required to observe C$_2$-loss from fullerenes on microsecond timescales \cite{chen07}. Formation of C$_{59}$ in a statistical process would involve similar if not higher internal energies, most of which would remain with the C$_{59}$ leading to further C-loss. For non-statistical knockout fragmentation, however, much of the collision energy is carried away by the knocked out C atom, leaving colder C$_{59}$. In the case of C$_{59}^-$ anions, the dissociation energy is comparable to the electron binding energy, and these decay channels can be expected to compete.

\subsection{Total destruction cross sections}
\label{sec_xsex}


\begin{figure}
 \includegraphics[width=0.98\columnwidth]{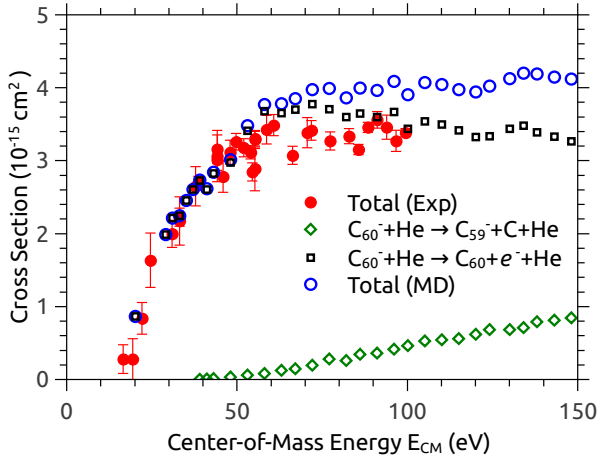}
 \caption{Absolute total destruction cross section for C$_{60}^-$ in collisions with He determined from beam attenuation measurements (filled circles). Carbon knockout cross sections (C$_{60}^-+$He$\rightarrow$C$_{59}^-$+C+He) from MD simulations, electron emission cross sections (C$_{60}^-+$He$\rightarrow$C$_{60}+e^-$+He) from the statistical model described in the text, and the sum of these two channels (open diamonds, squares, and circles, respectively).}
 \label{fig_totxsex}
\end{figure}

Experimental total destruction cross sections for C$_{60}^-+$He collisions are determined by measuring the attenuation of the C$_{60}^-$ beam as a function of He target density \cite{deRuette2018} and the results are shown for a range of different center-of-mass collision energies in Figure \ref{fig_totxsex}. Also shown in Figure \ref{fig_totxsex} are the single carbon knockout cross section from our MD simulations. These simulations indicate that knockout only gives a small contribution to the total C$_{60}^-$ destruction cross section. Furthermore, the simulations do not include secondary fragmentation or electron emission processes and should thus be considered upper limits for C$_{59}^-$ formation.

Our MD simulations model only the nuclear scattering part of the He-C$_{60}$ interaction and does not include the scattering of the He atom on the electrons in C$_{60}$ or delayed decay processes such as unimolecular dissociation (\textit{i.e.} statistical fragmentation) or thermionic electron emission. Depending on the kinetic energy of the incident C$_{60}^-$ ions, product ions travel for 10--40~$\mu$s between the gas cell and the mass analyzing system \cite{Stockett2014,deRuette2018}, giving ample time for such delayed processes to occur. Nevertheless, we can use the fullerene excitation energies from simulations to reproduce the measured total C$_{60}^-$ destruction cross section, which is presumably dominated by electron loss, using a statistical model. All parameters used in this model are taken from the present DFT calculation (Tables \ref{tab_eb} and \ref{tab_diss}) or from the literature as detailed below. None of these parameters are adjusted to obtain agreement with experiment. The same model will also be used in Section \ref{sec_td} to estimate the fraction of C$_{59}^-$ that is produced sufficiently cold to survive the flight from the gas cell to the end of the deflection field in the kinetic-energy-per-charge analyzer without losing the electron.

Here, we will first consider the survival probability of collisionally excited C$_{60}^-$ ions flying through the energy analyzer. First, the excitation energy deposited in the system is extracted from the MD simulations, taking the nuclear scattering on 60 carbon atoms into account. To this we add the internal energy of C$_{60}^-$ before the collision, which is taken to be 0.44~eV for $\sim$300~K C$_{60}$ as calculated by Yoo \textit{et al.} \cite{Yoo1992}. Finally, we add a small energy contribution from electronic stopping obtained by scaling the results by Schlath\"olter et al \cite{schlatholter99} to the collision velocity in the present experiment. Adding these three contributions, we arrive at a total internal energy $E_{int}$ for each individual trajectory in the MD-simulation, an internal temperature $T[K]=1000+(E_{int}[eV]-7.4)/C$ \cite{Andersen2001}. The effective temperature is then $T_{eff}=T-E_b/2C$, where $C=0.138$~K/eV is the heat capacity of C$_{60}$ \cite{Andersen2001}, and $E_b=2.664$~eV is the electron binding energy of C$_{60}^-$ (the electron affinity of C$_{60}$) \cite{Stochkel2013}. We calculate the survival probability for C$_{60}^-$ ions as a function of their internal energy $E_{int}$ by applying an Arrhenius expression for the electron detachment rate $k=\nu e^{-E_b/k_BT_{eff}}$ \cite{tomita02}. Here the pre-exponential factor $\nu$ is taken to be 10$^{13}$~s$^{-1}$ \cite{Andersen2000}, and $k_B$ is Boltzmann's constant. This model gives a maximum internal energy of 16--17~eV for C$_{60}^-$ to survive our experimental timescales of 10-40 $\mu$s, depending on the acceleration energy.

By combining the statistical model for electron loss and the MD simulations for knockout, we obtain the total C$_{60}^-$ destruction cross section (knockout plus electron loss) shown in Figure \ref{fig_totxsex}. The agreement with the corresponding measured quantity is satisfactory, validating our computational approach, and indicating that electron emission is indeed the dominant C$_{60}^-$ destruction mechanism in this energy range. In particular we see an onset of the experimental total C$_{60}^-$ destruction cross section around 16~eV which is consistent with the results from the statistical model - with center-of-mass collision energies below 16~eV, the system cannot be sufficiently heated to emit an electron on the experimental timescale. 

\subsection{Product distributions}


\begin{figure}
\includegraphics[width=0.98\columnwidth]{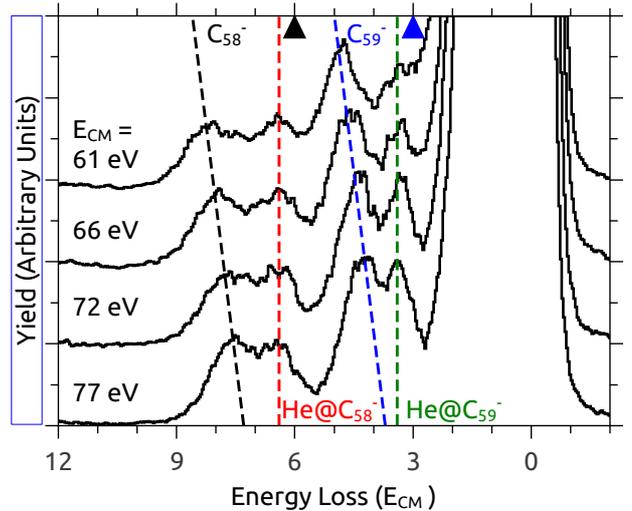}
\caption{Negatively charged product distributions following C$_{60}^-+$He collisions recorded at several values of $E_{CM}$. The dashed lines are to guide the eye. The measured energy loss (the difference in kinetic energy between the incident C$_{60}^-$ ion and the fragment anion is due to the combination of the change in mass and the energy transfer to the internal degrees of freedom of the molecular system and is given on the lower horizontal scale in units of $E_{CM}$. The expected kinetic energies of C$_{59}^-$ (blue) and C$_{58}^-$ (black) fragments with the same velocity as incident C$_{60}^-$ are indicated with arrowheads on the upper horizontal axis.}
\label{fig_spec}
\end{figure}

In Figure \ref{fig_spec} we show distributions of negatively charged reaction products of C$_{60}^-$+He collisions for center-of-mass collision energies ranging from $E_{CM}=61$~eV to $E_{CM}=77$~eV, which corresponds to kinetic energies of 11--14~keV keV for the incident C$_{60}^-$ ions in the laboratory reference frame. The collision products contributing to the energy spectra in Figure \ref{fig_spec} are those which are sufficiently cold to retain their negative charge through the analyzing system after the gas cell. The spectra are measured under single-collision conditions as has been confirmed by measuring the C$_{60}^-$ beam attenuation as a function of pressure in the gas cell. The distributions are plotted on an $E_{loss}=E_{parent}-E_{product}$ scale -- the difference in kinetic energy of the parent ion C$_{60}^-$ and the product ion -- in units of the center-of-mass collision energy $E_{CM}$. The measured kinetic energy loss is due not only to the difference in mass between the incident C$_{60}^-$ ion and the product anion but also to the energy transferred to the internal degrees of freedom of the molecular system. Capture of the He atom by the fullerene cage also reduces the kinetic energy. The energy loss due only to mass loss from C$_{60}^-$ without energy transfer would be $3N_{loss}^C\times E_{CM}$ where $N_{loss}^C$ is the number of C atoms lost. The expected positions of C$_{59}^-$ and C$_{58}^-$ fragments with the same velocity as the incident C$_{60}^-$ ions would thus be $E_{loss}=3\times E_{CM}$ and $E_{loss}=6\times E_{CM}$, respectively, and are indicated with arrowheads on the upper axis of Figure \ref{fig_spec}.

The energy spectra are dominated by intact C$_{60}^-$ ions at $E_{loss}=0$. Four daughter anion peaks are observed: two just below the expected C$_{59}^-$ energy ($E_{loss}=3\times E_{CM}$), and two just below the expected C$_{58}^-$ energy ($E_{loss}=6\times E_{CM}$). Based on comparisons with our MD simulations (\textit{vide infra}), we assign in both cases the higher energy peaks to endohedral complexes. The daughter ion peak with the highest kinetic energy (smallest energy loss) is the first observation of an endohedral defect fullerene complex with an odd number of C atoms, namely He@C$_{59}^⁻$. 

The observed C$_{58}^-$ and He@C$_{58}^-$ products are most likely due to secondary losses of loosely bound C atoms from C$_{59}^-$ and He@C$_{59}^-$, respectively. We have calculated the C$_{59}^-\rightarrow$C$_{58}^-+$C dissociation energy to be 4.54~eV (see Table \ref{tab_diss}), and that for the endohedral complex is likely comparable. Direct statistical dissociation C$_{60}^-\rightarrow$C$_{58}^-$+C$_2$ has a dissociation energy of 10.1~eV according to our calculations and is not competitive with electron emission (2.664 eV \cite{Stochkel2013}) and thus should not contribute significantly to the negative product spectrum. 


\begin{figure}
\centering
\includegraphics[width=0.79\columnwidth]{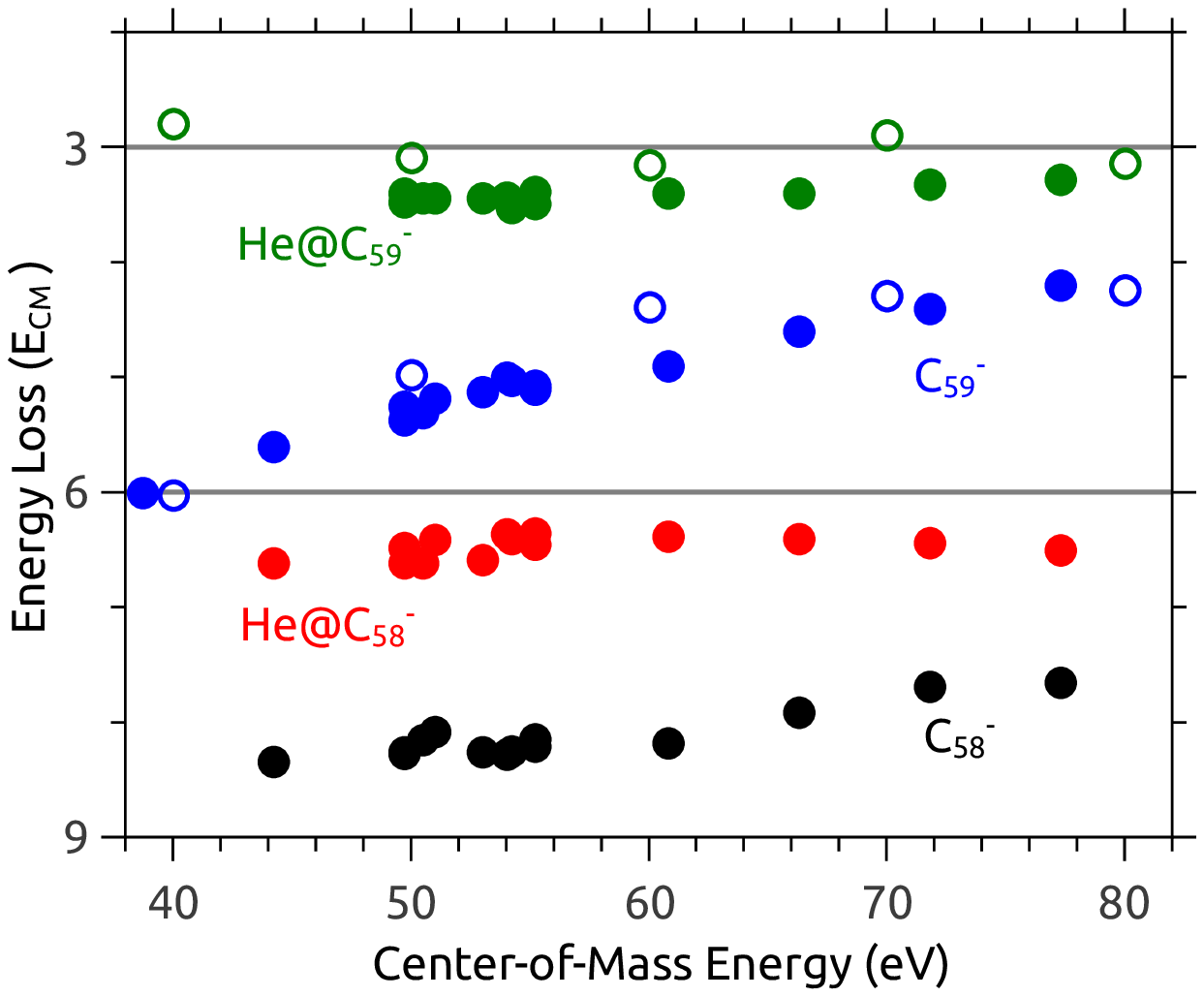}
\includegraphics[width=0.89\columnwidth]{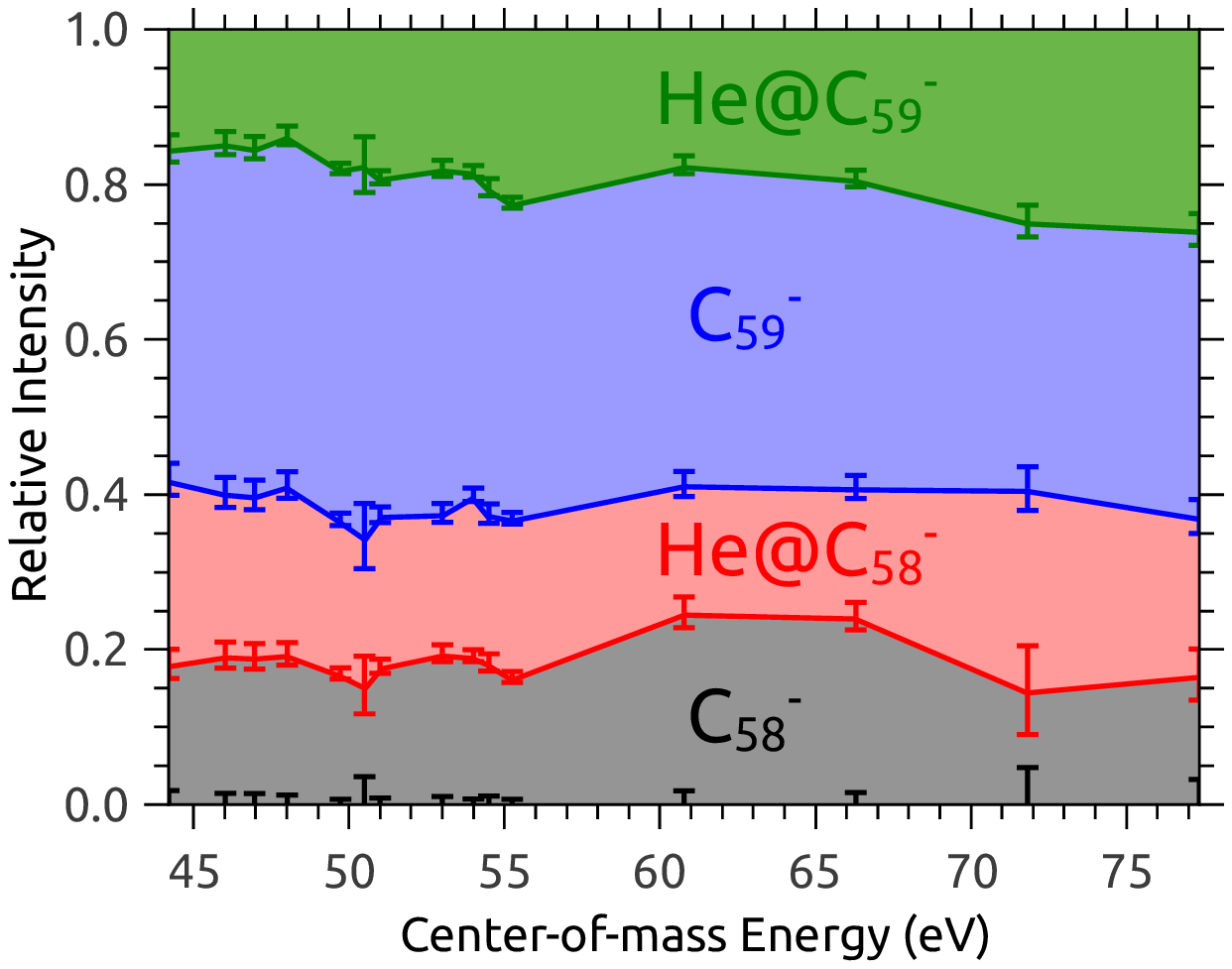}
\caption{Top: Mean energy losses of reaction products in units of $E_{CM}$; filled symbols: experiment, open symbols: MD simulations. The uncertainties in the mean position are smaller than the symbols. Bottom: Stacked area plot showing the relative intensities of the product peaks as a function of $E_{CM}$.}
\label{fig_trio}
\end{figure}

As can be seen in Figure \ref{fig_spec}, the peaks in the product distributions assigned to endohedrals remain roughly at the same $E_{loss}$ values, while the C$_{59}^-$ and C$_{58}^-$ peaks shift to greater $E_{loss}$ values with decreasing $E_{CM}$. In the top panel of Figure \ref{fig_trio}, we compare the measured and simulated mean energy loss for the observed product ions as a function of $E_{CM}$. For the experimental data, these values are extracted by fitting Gaussian peak shapes to each of the four daughter ions; two Gaussians are used for the tail of the parent C$_{60}^-$ peak. The peaks assigned to endohedral complexes are shifted by between 0.25 to 0.5 units of $E_{CM}$ relative to the expected $E_{loss}$ values for C$_{59}^-$ and C$_{58}^-$ with the same velocity as the incident C$_{60}^-$ ion at $3\times E_{CM}$ and $6\times E_{CM}$, respectively, which are indicated by horizontal lines in the upper panel of Figure \ref{fig_trio}. This is different from the situation when atoms like He are captured by intact C$_{60}$, where the increase in mass induces a well-defined energy loss equal to $E_{CM}$ \cite{caldwell92}. For He@C$_{59}^-$ formation, the knocked out carbon atom carries away some energy and energy losses smaller than $E_{CM}$ are therefore possible. Knockout collisions without capture giving C$_{59}^-$ and C$_{58}^-$ have a broader range of possible energy transfers and are thus shifted by larger energies on average. Mean energy losses from our MD simulations are in good agreement with the experimental values for He@C$_{59}^-$ and C$_{59}^-$. It is important to note that the MD simulations reproduce the different dependencies of the center-of-mass collision energy for both these fragments -- the energy loss increases with decreasing collision energy for the C$_{59}^-$ fragment while the energy loss of the endohedral complex He@C$_{59}^-$ remains close to constant in both the experiment and in the simulations.

The energy shifts of the experimentally observed He@C$_{58}^-$ and C$_{58}^-$ products relative to the expected energy of C$_{58}^-$ are similar to those of C$_{59}^-$ and He@C$_{59}^-$, respectively. This is consistent with our view that these products originate from C$_{59}^-\rightarrow$C$_{58}^-$+C as a second step after knockout. Our MD simulations do not include statistical fragmentation processes such as C$_{59}^-\rightarrow$C$_{58}^-$+C and we are thus unable to compare simulation with experiment in this case. 


In the bottom panel of Figure \ref{fig_trio}, one can see that the relative intensities of the four anion fragment peaks (again extracted using Gaussian fits) vary little as a function of the collision energy down to 45~eV, beyond which we cannot resolve He@C$_{59}^-$ from the tail of the C$_{60}^-$ peak. This supports our interpretation that all four products arise from a common mechanism with a single threshold energy.

\subsection{Threshold displacement energy}
\label{sec_td}

\begin{figure}
\includegraphics[width=0.98\columnwidth]{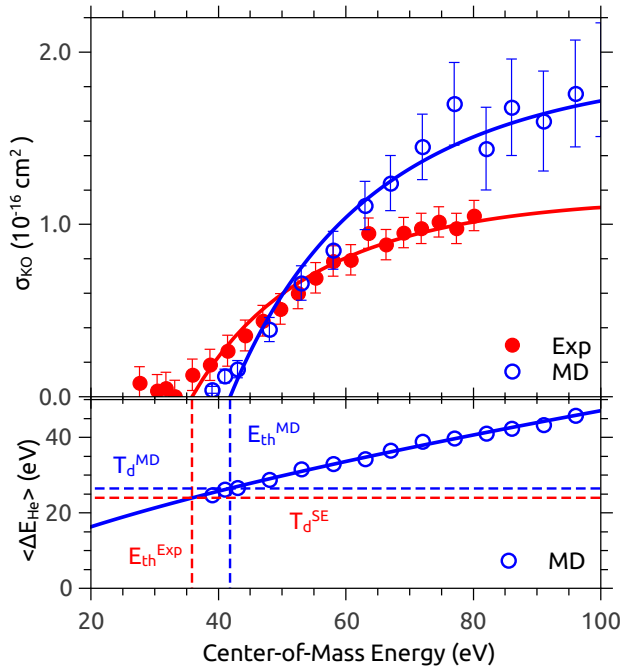}
\caption{Above: experimental and simulated cross sections for detection of daughter anions arising from non-statistical fragmentation in C$_{60}^-+$He collisions. For the simulations, C$_{59}^-$ daughters predicted to be lost to electron emission by our statistical model are excluded from the cross section (\textit{cf.} Figure \ref{fig_thresh}). The solid lines are fits to Eq. (\ref{eqn_tao}) which gives the threshold energy $E_{th}$. Below: MD-simulated mean He-C energy transfers $\left<\Delta E_{He} \right>$ in collisions leading to C$_{59}$ fragments cold enough to survive the experiment. The solid line is a power law fit; dashed lines illustrate the determination of the threshold displacement energy $T_d$.}
\label{fig_thresh}
\end{figure}

As discussed above, the four different reaction channels leading to the production of He@C$_{59}^-$,  C$_{59}^-$, He@C$_{58}^-$, or C$_{58}^-$ fragments are all non-statistical in nature and can be considered to be due to a common mechanism. Accordingly, we deduce the total absolute cross section for non-statistical fragmentation from our mass spectra by summing the intensities of all four anion fragment peaks and relating this sum to the C$_{60}^-$ total destruction cross section as described in Section \ref{sec_exp}. The resulting experimental cross section for C$_{60}^-$+He collisions yielding negatively charged fragments is given in the upper panel of Figure \ref{fig_thresh}.  

Also included in Figure \ref{fig_thresh} are results from our MD simulations combined with our statistical model for delayed electron emission from C$_{59}^-$ products. In the upper panel we give the cross section for single carbon knockout and in the lower panel the average energy transfer to the molecular system $\left< \Delta E_{He} \right>$ in collisions leading to C knockout. Included in both of these quantities are only those trajectories which leave the resulting C$_{59}$ sufficiently cold to survive the timescale of our experiment. To calculate the survival probability of C$_{59}^-$, we first determine the internal energy of C$_{59}$ at the end of each MD trajectory. The heat capacity for C$_{59}$ is found by scaling that of C$_{60}$ by the number of degrees of freedom, and we use a value of 3.56~eV for the electron binding energy (Table \ref{tab_eb}). Due to the higher electron binding energy of C$_{59}^-$ compared to C$_{60}^-$, our statistical model gives a higher internal energy cutoff of 24-25~eV for C$_{59}^-$ to survive until detection in the experiment. The effects of these considerations are negligible for $E_{CM}<60$~eV but there are significant corrections for higher energies. For comparison to experiment, we do not need to consider the competitive C$_{59}^-\rightarrow$C$_{58}^-$+C channel, as C$_{58}^-$ is included in our total non-statistical product cross section. 

The experimental and simulated cross sections are qualitatively similar, albeit with an offset in the observed threshold energy for knockout. Such an offset has been observed previously in comparison between experimental and simulated knockout cross sections \cite{Stockett2015}, and may be due to the approximate, classical nature of the Tersoff potential used to model the C-C bonds. To determine the threshold energy for carbon knockout, we model the experimental and MD cross section assuming that the primary process is an elastic binary collision between a He projectile with kinetic energy equal to the He-C$_{60}^-$ center-of-mass $E_{CM}$ and a free C atom at rest. This is justified in that the knockout process happens on an ultrafast timescale on which the remaining 59 C atoms are standing still. For such collisions, Chen \textit{et al.} \cite{Chen2014} give the following expression, based on Lindhard scattering theory \cite{Lindhard1968a}, for the cross section $\sigma_{KO}$ leading to energy transfer above a given fixed value (here the displacement energy $T_d$):

\begin{equation}
\sigma_{KO}=\frac{A/E_{CM}}{\pi^2 \arccos^{-2} (\sqrt{E_{th}/E_{CM}})-4},
\label{eqn_tao}
\end{equation}

\noindent where $E_{th}$ is the minimum center-of-mass energy (for the He-C$_{60}$ system) required to transfer $T_d$. We take $A$ and $E_{th}$ as fit parameters, yielding $E_{th}^{exp}=35.8\pm$0.5~eV and $E_{th}^{MD}=41.8\pm 1.5$~eV. As seen in Figure \ref{fig_thresh}, this simple model accurately reproduces both the experimental and MD cross sections above $E_{th}$. There is some (barely significant) deviation close to threshold, which probably due to the fact that $T_d$ is not single-valued, but rather varies somewhat with respect to the angles between the imparted momentum and the molecular bonds \cite{cui94,Gatchell2016}, as well as with the stretching of the bonds due to molecular vibrations \cite{Meyer2012}.

The threshold energy $E_{th}$ is projectile dependent \cite{postma14}; to obtain the intrinsic threshold displacement energy $T_d$ we need the energy transferred from the projectile (He) to the target at threshold, which we extract from our MD simulations. From a power-law fit to $\left< \Delta E_{He} \right>=c\times {E_{CM}}^P$ we obtain the mean energy transfer at the experimental knockout threshold $E_{CM}=E_{th}^{exp}$. This is our semi-empirical value for the C$_{60}\rightarrow$ C$_{59}$+C threshold displacement energy, $T_d^{SE}= 24.1 \pm 0.5$~eV. The uncertainty given here is calculated from the uncertainty in the power-law fit parameters $c$ and $P$, and $E_{th}^{exp}$. Our semi-empirical value of $T_d$ is much higher than generally assumed previously for fullerenes (around 15~eV \cite{Fueller1996,Parilis1994,Larsen1999,Zettergren2013}), and is similar to those measured for the planar $sp^2-$hybridized carbon systems graphene (23.6~eV \cite{Meyer2012,Meyer2013}) and PAHs ($23.3\pm 0.3$~eV \cite{Stockett2015}). 

As our MD simulations give a somewhat higher threshold energy than the measurements we find a correspondingly higher threshold displacement energy $T_d^{MD}=26.5\pm 0.8$~eV, which is slightly lower than previous simulation-based results \cite{cui94}.

\begin{figure}
\includegraphics[width=0.98\columnwidth]{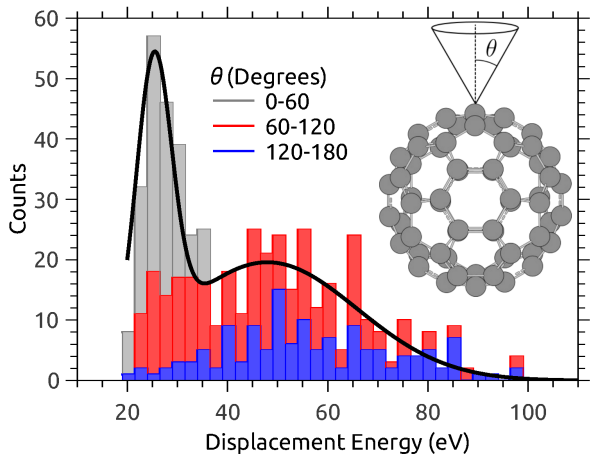}
\caption{Histogram of displacement energies from projectile-free MD simulations. The solid line is a fit to the sum of two Gaussians describing a bimodal distribution.}
\label{fig_bimodal}
\end{figure}

We have also performed projectile-free MD simulations, where an individual C atom is simply assigned a certain velocity at a given angle relative to the axis connecting the atom and the center of the fullerene cage. The minimum energy at which the atom is removed is the displacement energy for that angle. A histogram of these results is shown in Figure \ref{fig_bimodal}, which shows a clear bimodal distribution. A fit to the sum of two Gaussians gives the centroid of the low-energy mode as 25.3$\pm$0.4~eV with a width of 7$\pm$1~eV. This is consistent with our result for the full simulations including the He projectile. Trajectories contributing to the high-energy mode of the histogram in Figure \ref{fig_bimodal} would presumably be left with internal energies too high to survive the experiment.

Notably, most of the trajectories contributing to the low-energy mode are at angles pointed outwards from the center of the fullerene cage. This helps explain the surprisingly high abundance of endohedral complexes in our experiments. The He projectile must first penetrate the cage prior to the knockout collision, in which a large fraction of its energy is transferred to the knocked out C atom, leaving the backscattered He with insufficient energy to escape the (now defective) cage.

\section{Conclusions}

This is, to our knowledge, the first determination of a threshold displacement energy for free fullerenes or for any type of fullerene based material. This intrinsic material property is a key parameter for modeling radiation damage in many contexts, for example during electron microscopy imaging \cite{Meyer2012} or gas-phase reactions in the interstellar medium \cite{postma14}. Finally, the surprising observation of the endohedral defect fullerene complex He@C$_{59}^-$ is a remarkable testament to the intriguing complexity of fullerene reactions, which continue to fascinate more than 30 years after their discovery. 

\section{Acknowledgments}
This work was performed at the Swedish National Infrastructure, DESIREE (Swedish Research Council Contract No.\ 2017-00621). It was further supported by the Swedish Research Council (grant numbers 2014-4501, 2015-04990, 2016-03675, 2016-04181, 2016-06625). See Supplemental Material for videos of selected MD simulations.

\end{document}